\documentstyle[aps,prl]{revtex}

\begin{document}
\draft

 %Inserted by TeXtelmExtel

\twocolumn[\hsize\textwidth\columnwidth\hsize\csname @twocolumnfalse\endcsname

 %Inserted by TeXtelmExtel

\title{$^{89}$Y NMR imaging of the staggered magnetization in doped
Haldane chain Y$_2$BaNi$_{1-x}$Mg$_x$O$_5$}

 %Inserted by TeXtelmExtel

\author{ F. Tedoldi and R. Santachiara} 
\address{
Department of Physics "A. Volta", 
Unita' INFM di Pavia-Via Bassi, 6 - 27100-I Pavia, Italy}

 %Inserted by TeXtelmExtel

\author{M. Horvati\'{c}}
\address{
Grenoble  High  Magnetic  Field  Laboratory,  CNRS and MPI-FKF, 
B.P. 166, 38042 Grenoble Cedex 09, France}

 %Inserted by TeXtelmExtel

\date{\today}
\maketitle
\widetext
\begin{abstract}

 %Inserted by TeXtelmExtel

We present a {\it static}, model-independent, experimental 
determination of the spin-spin correlation length $\xi$ in a quantum 
spin system. This is achieved in doped Haldane (i.e., $S=1$ Heisenberg 
antiferromagnetic) chain Y$_2$BaNi$_{1-x}$Mg$_x$O$_5$ by 
$^{89}$Y NMR imaging of the staggered magnetization induced around the 
Mg-impurities (i.e., chain boundaries) by magnetic field. The magnitude 
of this magnetization is found to decay exponentially with $\xi$ equal 
to the theoretical prediction for an infinite $S=1$ chain, and the staggered
magnetic moment at the edge site showing the Curie behavior of an effective 
$S=1/2$ spin.

 %Inserted by TeXtelmExtel

\end{abstract}
\pacs {PACS numbers: 75.10.Jm, 76.60.-k}
] \newpage %%%%%%%%%%%%%%%%%%%%
% \leftskip 54.8pt
% \rightskip 54.8pt
%%%%%%%%%%%%%%%%%%%%

 %Inserted by TeXtelmExtel

%%%%%%%%%%%%%%%%%%%%
\narrowtext
%%%%%%%%%%%%%%%%%%%%
\bigskip One dimensional (1D) quantum antiferromagnetic (AF) spin systems
such as spin-ladders, spin-Peierls systems and spin chains have recently
attracted an increasing interest, partly as a possible way to approach from
a lower dimension the physics of the copper-oxide high $T_{c}$\
superconductors. In all these systems the spin-spin correlation function is
a quantity of fundamental importance, which can be experimentally probed as
a function of wave vector and energy by inelastic neutron scattering, or by
studying the {\it static} response of the system to the perturbation created
by impurities, through NMR imaging of local spin polarization. In recent
examples of this latter technique applied to high $T_{c}$\ superconductors 
\cite{Bobroff}, spin-ladder systems \cite{bib02} or $S=1/2$ chains \cite
{bib01}, the broadening of the NMR lines (or the modification of the
lineshape) due to impurities has always been interpreted within a specific
model in order to obtain information on the correlation length $\xi $,
making conclusions manifestly {\it model dependent}. In this letter we
present an NMR study of a doped Haldane ({\it i.e.}, $S=1$ Heisenberg AF
spin) chain, in which individual spin sites near impurities are resolved as
satellite peaks in the NMR spectra. In this way the real-space dependence of
the spin polarization is followed site by site, providing a complete and
model independent picture of the impurity screening, and giving thus a
direct access to $\xi $. Such a study should be compared to the observation
of satellite lines revealing screening of normal \cite{Friedel} or Kondo 
\cite{Boyce} impurities in a metal, but it applies here to a low-dimensional
quantum antiferromagnet, {\it i.e.}, to a class of systems in which
screening of non-magnetic impurities has generated recently a large body of
theoretical works \cite{Dagotto}.

 %Inserted by TeXtelmExtel

The main characteristic of integer spin chains is the existence of a gap
between the lowest energy levels, predicted by Haldane \cite{bib1} and
experimentally observed in several compounds \cite
{bib2,bib3,bib4,bib5,bib6,bib7}. If such a system is doped by non-magnetic
impurities, an ideal infinite chain is ''cut'' to an ensemble of open
segments of length $L$, in which the lowest lying eigenstates are a singlet $%
|0\rangle $ and a triplet $|1,S_{T}^{z}\rangle $, with an energy separation
exponentially decreasing with $L$ \cite{bib8}. The Haldane gap, \cite{bib1}
defined for $L\rightarrow \infty $, is the one above the resulting fourfold
quasi-degenerate ground state. In the state $|1,1\rangle $ it has been shown
by numerical calculations \cite{bib10,bib11,bib12,bib13}, that the
expectation value of the local spin operator $\langle S_{i}^{z}\rangle $
decreases exponentially \cite{cmt1} from the boundary with staggered phase.
In particular, for a pure Heisenberg model, one has an effective spin-1/2 at
the edge of the chain ($\langle S_{1}^{z}\rangle \simeq 1/2$) and a
characteristic decay length $\xi \simeq 6$ \cite{bib10}, a value very close
to the zero-temperature correlation length ($\xi _{\infty }$) of an infinite
chain. Thus, it has been argued that the zero-temperature local
magnetization (corresponding for long chains to $\langle S_{i}^{z}\rangle $
in the state $|1,1\rangle $) reflects indeed the decay of the correlation
function in the Haldane phase.

 %Inserted by TeXtelmExtel

$S=1/2$ boundary spins have been experimentally observed in low-temperature
ESR spectra of pure and doped Ni(C$_{2}$H$_{8}$N$_{2}$)$_{2}$NO$_{2}$(ClO$%
_{4}$) \cite{bib14,bib15} and Y$_{2}$BaNiO$_{5}$ \cite{bib16}. Also
specific heat measurements \cite{bib17} can be explained in terms of $S=1/2$
excitations \cite{bib18}. On the other hand, no direct experimental
determination of the site dependence of the local magnetization have been
reported until now. Moreover, the temperature dependence of the expectation
value of edge spins, and the relationship between the decay of the staggered
magnetization and the infinite chain correlation function have not been
established yet at temperatures where the excited states are populated.

 %Inserted by TeXtelmExtel

The compound Y$_{2}$BaNiO$_{5}$, due to the low inter/intrachain coupling
ratio ($|J^{\prime }/J|\leq 5\times 10^{-4}$, with $J=285$ K) and the small
single-ion anisotropy \cite{bib5}, is a prototype one-dimensional $S=1$
Heisenberg AF. Moreover, a controlled substitution of Ni$^{2+}(S=1)$ by Mg$%
^{2+}(S=0)$ offers the possibility to obtain open chains of the average
length $\langle L\rangle \cong 1/x$. In Y$_{2}$BaNi$_{1-x}$Mg$_{x}$O$_{5}$,
Ni$^{2+}$ local magnetization generates a site dependent hyperfine field at $%
^{89}$Y nuclei inducing a multi-peak structure in $^{89}$Y NMR spectra.
These spectra are sort of a map of $\langle S_{i}^{z}\rangle $, inaccessible
by any other technique because a non-homogeneous, variable-amplitude
response to the external field is involved.

 %Inserted by TeXtelmExtel

$^{89}$Y NMR measurements have been carried out on $x=0.05$ and $x=0.10$
powder samples of Y$_{2}$BaNi$_{1-x}$Mg$_{x}$O$_{5}$, at the NMR facility of
the Grenoble High Magnetic Field Laboratory. The $^{89}$Y NMR spin-echo
spectra have been obtained at fixed frequency by sweeping the magnetic
field. Fig. 1 shows representative spectra for the $x=0.05$ sample. Several
peaks can be observed, their shift from the central line increasing with
decreasing temperature. In the following we show that this temperature
dependence yields directly the temperature evolution of the local
magnetization. The shift $\,\Delta (p_{I})=|H_{0}(p_{I})-H_{0}(p_{c})|\,$ of
the edge-spin peak $p_{I=1}$ with respect to the central peak $p_{c}$ is
reported in Fig. 2 as a function of $T$. A comparison of spectra for two
samples with different average chain lengths shows that for $T\simeq J$ the
peak positions are independent on $\langle L \rangle $ (see the inset to
Fig. 2). Fig. 3a shows the position dependence $\,\Delta (p_{I})$ at
different temperatures. It is found that the shift decreases exponentially
with $I$ at all temperatures above $\sim 100$ K. Below 100~K the peaks are
smeared into a single broad line (inset), preventing us to extend the
analysis to lower temperature.

 %Inserted by TeXtelmExtel

The intensity of the NMR spectrum is proportional to the number of $^{89}$Y
nuclei which, for the irradiation frequency $\nu _{RF}$, obey the resonance
condition 
\begin{equation}
\nu _{RF}={\frac{{\gamma }}{{2\pi }}}(H_{0}+h_{j}^{z})={\frac{{\gamma }}{{%
2\pi }}}\left[ \;H_{0}+\left( \sum_{l,k}\hat{A}_{lkj}\,\langle \vec{S}%
_{lk}\rangle \right) ^{z}\;\right]
\end{equation}
where $({\gamma }/{2\pi })=2.0859MHz/Tesla$ is the $^{89}$Y gyromagnetic
ratio, ${h}_{j}^{z}$ the $z$-component of the hyperfine field at the $j$-th
nucleus and $\hat{A}_{lkj}$ the Ni$^{2+}$-$^{89}$Y hyperfine coupling
tensor. In Eq. (1) $k$ is a chain index and $l$ is an in-chain site index. A
given $^{89}$Y has two Ni$^{2+}$ nearest neighbors ($nn$) belonging to the
same chain and two next nearest neighbors ($nnn$) on adjacent chains \cite
{bib19}. On the basis of the NMR results in pure Y$_{2}$BaNiO$_{5}$ one can
assume that only $nnn$ sites contribute to ${h}_{j}^{z}$ : 
\begin{eqnarray}
A_{lkj}^{\alpha \beta } &=&A=\text{13\ kG}\;\;\;\;%
\mbox{if $\alpha = \beta$ and ($j$) -
$(l,k)$ are $nnn$}  \nonumber \\
A_{lkj}^{\alpha \beta } &=&0\hspace{1.7cm}\;%
\mbox{if $\alpha \neq \beta$ or
($j$) - $(l,k)$ are not $nnn$}  \nonumber \\
&&
\end{eqnarray}

 %Inserted by TeXtelmExtel

In the following we summarize the main reasons supporting this assumption 
\cite{bib20}. The NMR powder spectra \cite{bib7} having a shift much larger
than the line broadening indicate that the hyperfine coupling is essentially
scalar and hence due to a transferred hyperfine interaction. Since the
mechanism of transferred hyperfine interaction requires extension of the Ni$%
^{2+}-3d$ orbitals at the Y site, only $nn$ and $nnn$ Ni$^{2+}$can couple to 
$^{89}$Y. On the other hand, as already noticed \cite{bib6}, the lattice
position of the $^{89}$Y is strongly unfavorable for transferred interaction
from $nn$ Ni$^{2+}$, and the coupling can be assumed essentially limited to
two equivalent contributions from $nnn$ ions only. Since for an infinite
chain in uniform magnetic field the magnetization is uniform, $\langle \vec{%
S_{lk}}\rangle =\langle \vec{S}\rangle $, from the standard plot of the
shift of the NMR line as a function of the spin susceptibility one derives $%
\sum_{l,k}{A_{lkj}}=2A=$ 26\ kG. Furthermore, small interchain coupling $%
J^{\prime }$ allows us to consider the $nnn$ ions (which belong to different
chains) contributing to ${h}_{j}^{z}$\ as completely uncorrelated. Thus,
Eqs. (1) and (2) link the position of the maxima in the NMR spectra to the
different values of the local magnetization along one chain. In particular,
the high temperature spectra shown in Fig. 1 correspond to a large number of
Ni$^{2+}$ spins having the same expectation value $\langle S_{i}^{z}\rangle
\equiv \langle S_{c}^{z}\rangle $ (central peak $p_{c}$), and to a certain
number of sites for which $\langle S_{i}^{z}\rangle \neq \langle
S_{c}^{z}\rangle $, yielding the peaks $p_{I}$. According to $T=0$ numerical
results \cite{bib10,bib11,bib12,bib13}, one expects $|\langle
S_{i}^{z}\rangle -\langle S_{c}^{z}\rangle |\rightarrow 0$ for $i\rightarrow
L/2$. Thus we map the peaks $p_{I}$ with the positions $I$ and $(L-I+1)$
along the chain, e.g., positions $1$ and $L$ correspond to $p_{1}$.

 %Inserted by TeXtelmExtel

As shown in Fig. 2 (solid line) the temperature evolution of $\,\Delta
(p_{1})$ is very well accounted for by a hyperfine field of the form 
\begin{equation}
\lbrack \Delta (p_{1})](T)=A\,{\frac{g\mu _{B}S(S+1)}{3k_{B}T}}%
\,H_{0}
\end{equation}
with $S=1/2$, $g=2$ and $H_{0}=$ 14~Tesla. This means that, for $\Delta
_{H}\lesssim T\lesssim J$ (where $\Delta _{H}\approx 100$\thinspace K is the
Haldane gap for pure Y$_{2}$BaNiO$_{5}$), the non-uniform magnetization
associated to the extreme sites of the chain is close to the one
corresponding to ''free'' $S=1/2$ spins. Since no doping ({\it i.e.}, $%
\langle L\rangle $-) dependence of $\Delta (p_{1})$ have been found at room
temperature (inset to Fig. 2), this conclusion is independent on the
chain-length at least for $\langle L\rangle \gtrsim 10$ and $T\simeq J$. One
could observe that a $S=1/2$ Curie-like behavior can be obtained as thermal
expectation value of the edge spins on the lowest energy states (singlet and
triplet) of the {\it finite} chain, provided they are separated by an energy
gap $\Delta E<<g\mu _{B}H_{0}$. According to Monte Carlo calculations \cite
{bib10}, $\Delta E$ is indeed considerably smaller than the Zeeman energy in
a field of 14 T ($\simeq \ 0.08J$) for $L\simeq 20$ ({\it i.e.,} $x=0.05$),
while $\Delta E\approx g\mu _{B}H_{0}$ for $L\simeq 10$ ({\it i.e.,} $x=0.10$%
). However, one has to notice that for $T>\Delta _{H}$ a precise calculation
of the thermodynamic properties of the system has to take into consideration
also the excited states, as is evident from the analysis of the spin-spin
correlation reported below.

 %Inserted by TeXtelmExtel

The dependence $\Delta (p_{I})$ on $I$ can be fitted to an exponential decay 
\begin{equation}
\lbrack \Delta (p_{I})](T)=[\Delta (p_{1})](T)\,e^{-(I-1)/\xi (T)}\;
\end{equation}
(solid lines in Fig. 3a), and the characteristic decay length $\xi (T)$ of
the magnetization can been extracted. As shown in Fig. 3b, $\xi (T)$
obtained in this way is very close to the values of the {\it infinite} chain
correlation length $\xi _{\infty }$ calculated by Kim $et~al.$ \cite{bib21}.
This remarkable conclusion indicates that the picture of boundary defects,
which reflects the bulk correlation of an infinite chain, holds not only in
the Haldane phase ($T<<\Delta _{H}$) \cite{bib13}, but in general for $T<J$.
A possible way to interpret this result can be based on a heuristic
extension from zero to finite temperature of the model proposed in Ref.\cite
{bib12}. There the expectation values of $S_{i}^{z}$ in the state $%
|1,1\rangle $ are shown to decay exponentially over a length $\xi =\xi
_{\infty }(T=0)$, due to the interaction of the fields describing the
translational invariant chain in the free boson model, with the effective
fields originating from two end of chain $S=1/2$ spins. The experimental
findings reported in Fig. 3 are then immediately recovered if one assumes
that these edge ''extra'' spins are practically Curie-like (as indicated by
the behavior of $p_{1}$ in Fig. 2), and substitutes $\xi _{\infty }(T=0)$ by 
$\xi _{\infty }(T)$ and the expectation values of $S_{i}^{z}$ in $%
|1,1\rangle $ by the thermal average.

 %Inserted by TeXtelmExtel

At low temperature the progressive increase of the width $d(p_{I})$ of the
NMR peaks (see Fig. 1) prevents a meaningful analysis of the boundary
magnetization along the lines used for $T\gtrsim 100$ K. However, the
characteristic shape of the NMR spectra \cite{bib20} clearly indicates the
persistence of field-induced staggered magnetization. The origin of the
broadening is likely related to the anisotropy terms in the magnetic
Hamiltonian \cite{bib5}, which in powder samples causes a distribution of $%
\langle S_{i}^{z}\rangle $. This observation is corroborated by the fact
that the ratio between the easy-axis anisotropy $D=$\ 8 K and the Zeeman
energy is of the same order of magnitude as $d(p_{I})/\Delta (p_{I})$. In
order to extend the data to lower temperature it is thus necessary to work
on single crystals. However, an additional source of broadening is a
distribution of chain lengths around the average value $\langle L\rangle $,
which is expected to be important at $T\lesssim 100$ K. Indeed, at low
temperature $\xi $ is increasing, and for short chains with $L/2\lesssim \xi 
$ magnetization is becoming $L$-dependent. In principle these effects can be
resolved by decreasing the dopant concentration $x$, however, this will
proportionally decrease the size of our satellite peaks, {\it i.e.}, the
''resolution'' of NMR spectrum.

 %Inserted by TeXtelmExtel

In summary, through $^{89}$Y NMR in Mg doped Y$_{2}$BaNiO$_{5}$, we have
been able to image the local magnetization induced by ''free'' boundaries in a 
$S=1$ Heisenberg AF spin chain, in the temperature range $\Delta
_{H}\lesssim T\lesssim J$. By mapping the lattice positions onto the
different peaks in the NMR spectra, we have shown that the characteristic
decay length $\xi (T)$ of the boundary magnetization is equal to the
theoretical prediction for the correlation length of the Haldane chain. The
non-uniform magnetization of the edge spin in a finite chain follows a $S=1/2
$ Curie-like behavior. These results provide direct experimental
confirmation that the NMR of doped quantum AF spin systems is the right
technique to access the spin-spin correlation function. Improving the choice
of experimental system ({\it e.g.}, by using single crystals) and using
an appropriate nucleus for NMR imaging, this technique should provide more
reference data for comparison to the theory.

 %Inserted by TeXtelmExtel

The authors thank A. Rigamonti, P. Carretta, M.-H. Julien, P. Verrucchi, C.
Berthier and T. Feher for helpful discussions. Y$_{2}$BaNi$_{1-x}$Mg$_{x}$O$%
_{5}$ powder samples have been prepared at the Physical Chemistry Department
of the University of Pavia by M. Bini, D. Capsoni and V. Massarotti, who are
gratefully acknowledged. The research was supported by the CEE - TMR Program
under contract ERBFMGECT950077. Grenoble High Magnetic Field Laboratory is
Laboratoire conventionn\'{e} aux Universit\'{e}s J. Fourier et INPG Grenoble
I.

 %Inserted by TeXtelmExtel

 %Inserted by TeXtelmExtel

\begin{figure}[tbp]
\caption{$^{89}$Y NMR spectra in Y$_{2}$BaNi$_{0.95}$Mg$_{0.05}$O$_{5}$
recorded at fixed frequency $\protect\nu _{RF}=$ 29.4~MHz by sweeping
magnetic field. Resolved satellite peaks are labelled with the index $I$,
following the decreasing magnitude of their shift (measured from the central
line). In the inset, all the peaks are shown to be smeared in a single wide
line when the temperature is lowered.}
\end{figure}

 %Inserted by TeXtelmExtel

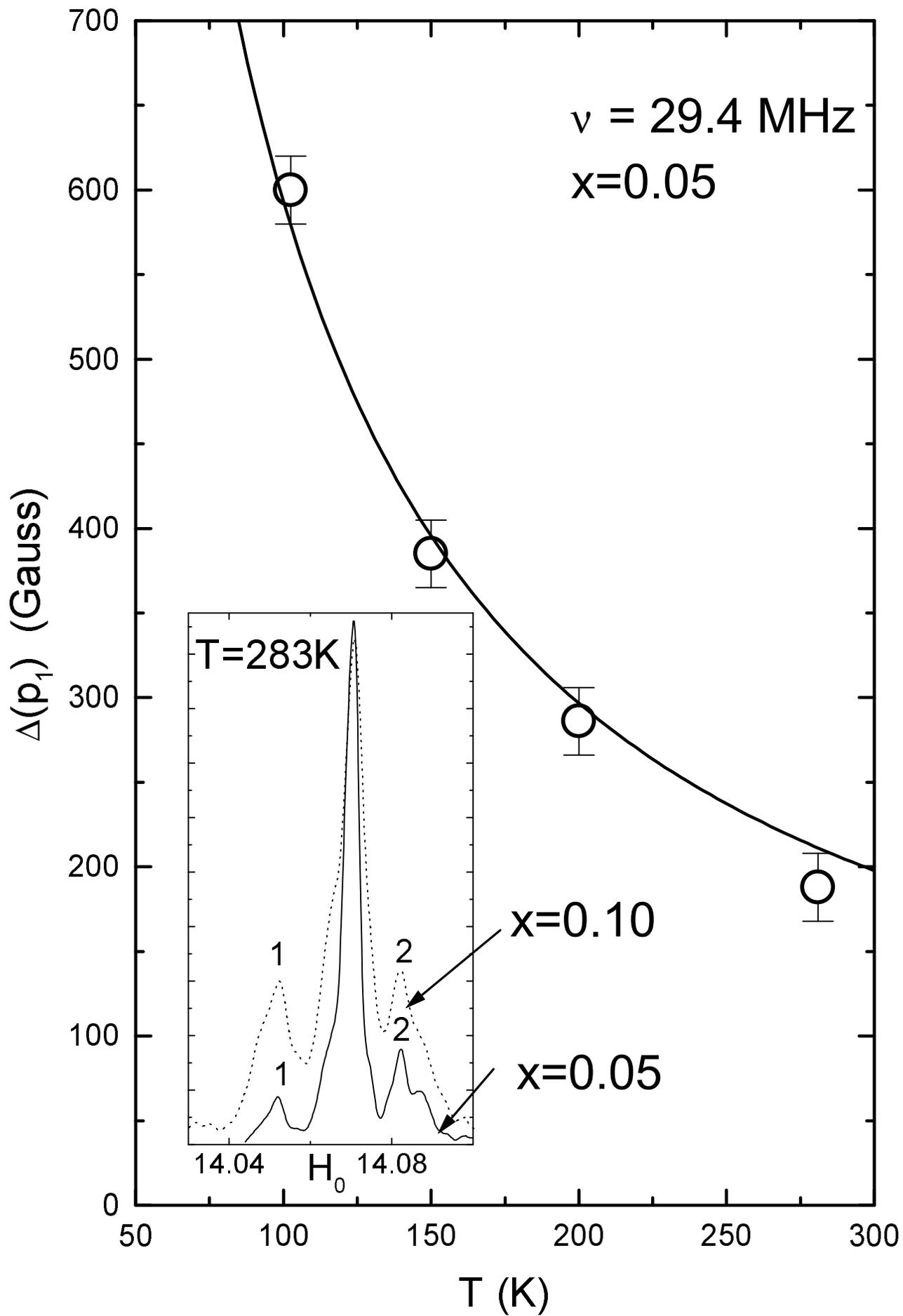
\begin{figure}[tbp]
\caption{Temperature dependence of the shift of the $p_{1}$ peak in $^{89}$Y
NMR spectra (measured from the central line, see Fig. 1). The solid line is
a Curie function for $S=1/2$ spins (Eq. 4 in the text). Inset shows the
doping dependence of the room temperature spectra of Y$_{2}$BaNi$_{1-x}$Mg$%
_{x}$O$_{5}$ demonstrating that peak positions do not depend on the chain
length.}
\end{figure}

 %Inserted by TeXtelmExtel

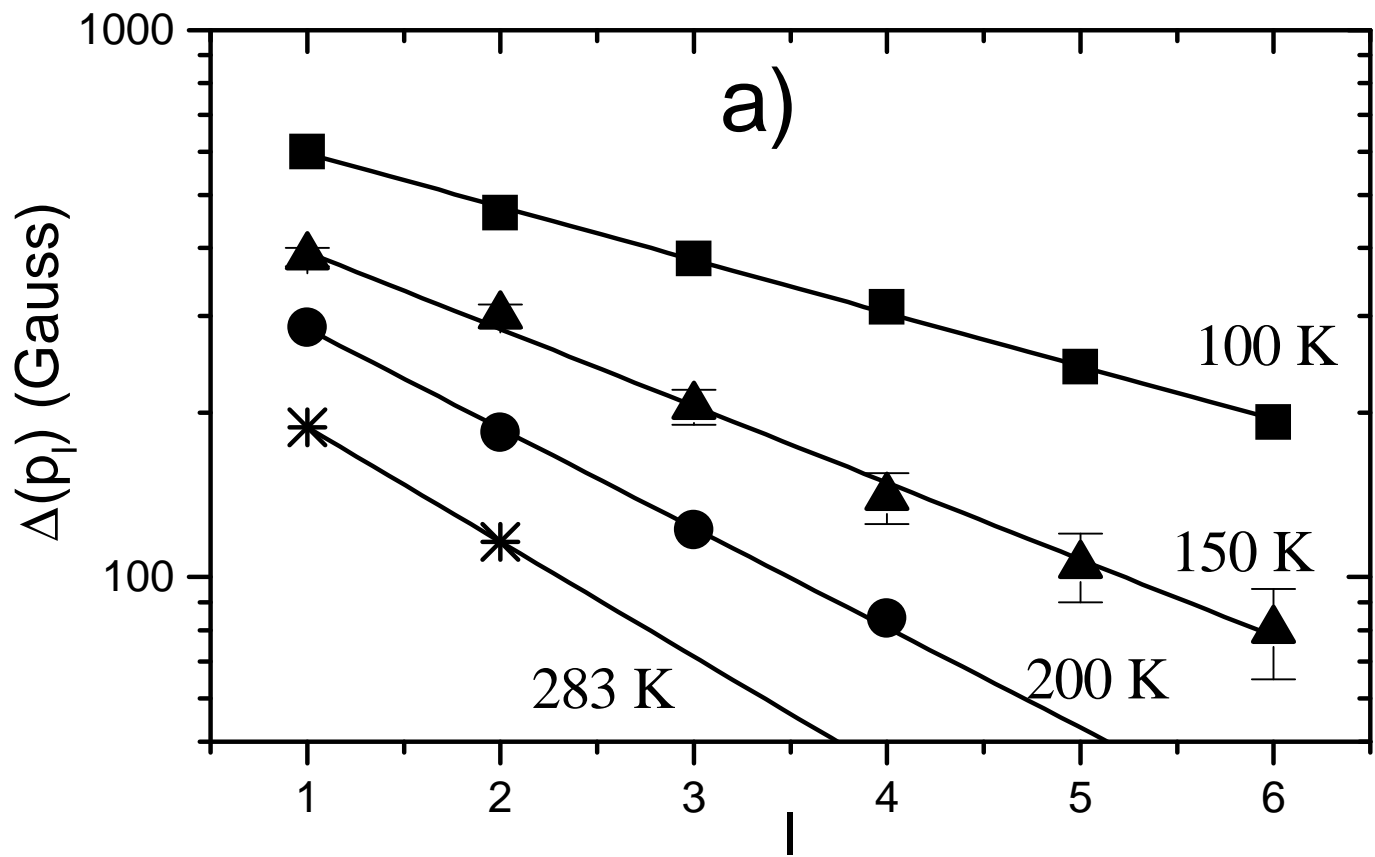
\begin{figure}[tbp]
\caption{a) Site-dependence of the magnitude of shift of $p_{I}$ peaks in Y$%
_{2}$BaNi$_{0.95}$Mg$_{0.05}$O$_{5}$ taken at $\protect\nu _{RF}=$ 29.4~MHz.
Solid lines are fits (Eq. 5 in the text), defining a decay length $\protect%
\xi (T)$ of the boundary magentization. b) Comparison of an experimental $%
\protect\xi (T)$ value (full squares) and the infinite chain correlation
length $\protect\xi _{\infty }(T)$ (up triangles) as calculated in Ref. 
\protect\cite{bib21}}
\end{figure}

 %Inserted by TeXtelmExtel


\begin{references}
\bibitem{cmt1}  In fact the ''exponential'' decay in the $T=0$ numerical
results is an interpolation, with the local magnitude $|\langle
S_{i}^{z}\rangle |$ for the odd (even) spin positions somewhat bigger
(smaller) than the interpolated exponential, ensuring that the total spin
integrated near each end of the chain is indeed \ $\sum_{i}\langle
S_{i}^{z}\rangle =1/2$. In our experimental data (Fig. 3) this ''fine
structure'' is not visible, probably because it is washed out by the
statistical average over the distribution of chain lengths $L$ and/or by
relatively high temperature.
\end{references}
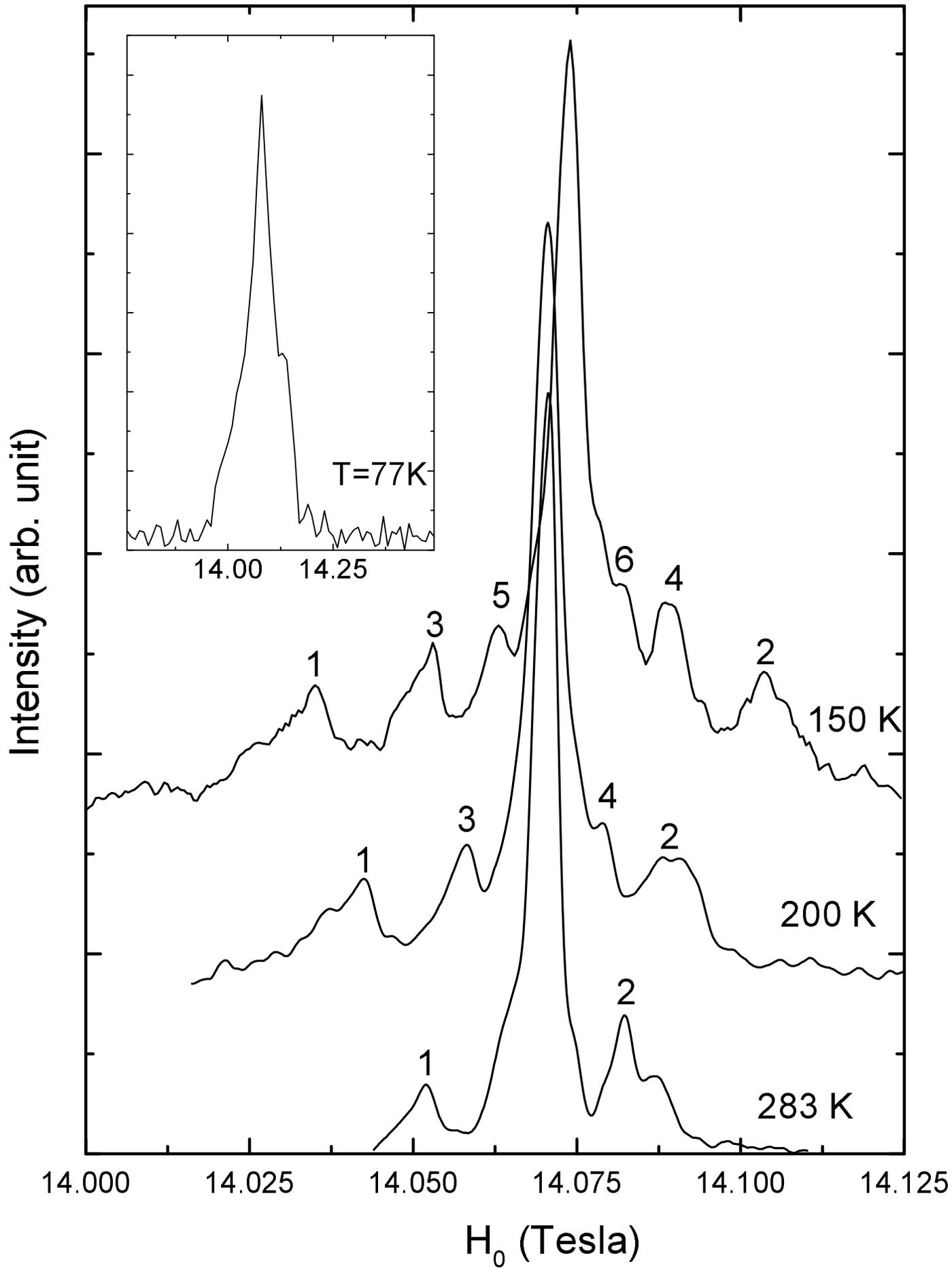
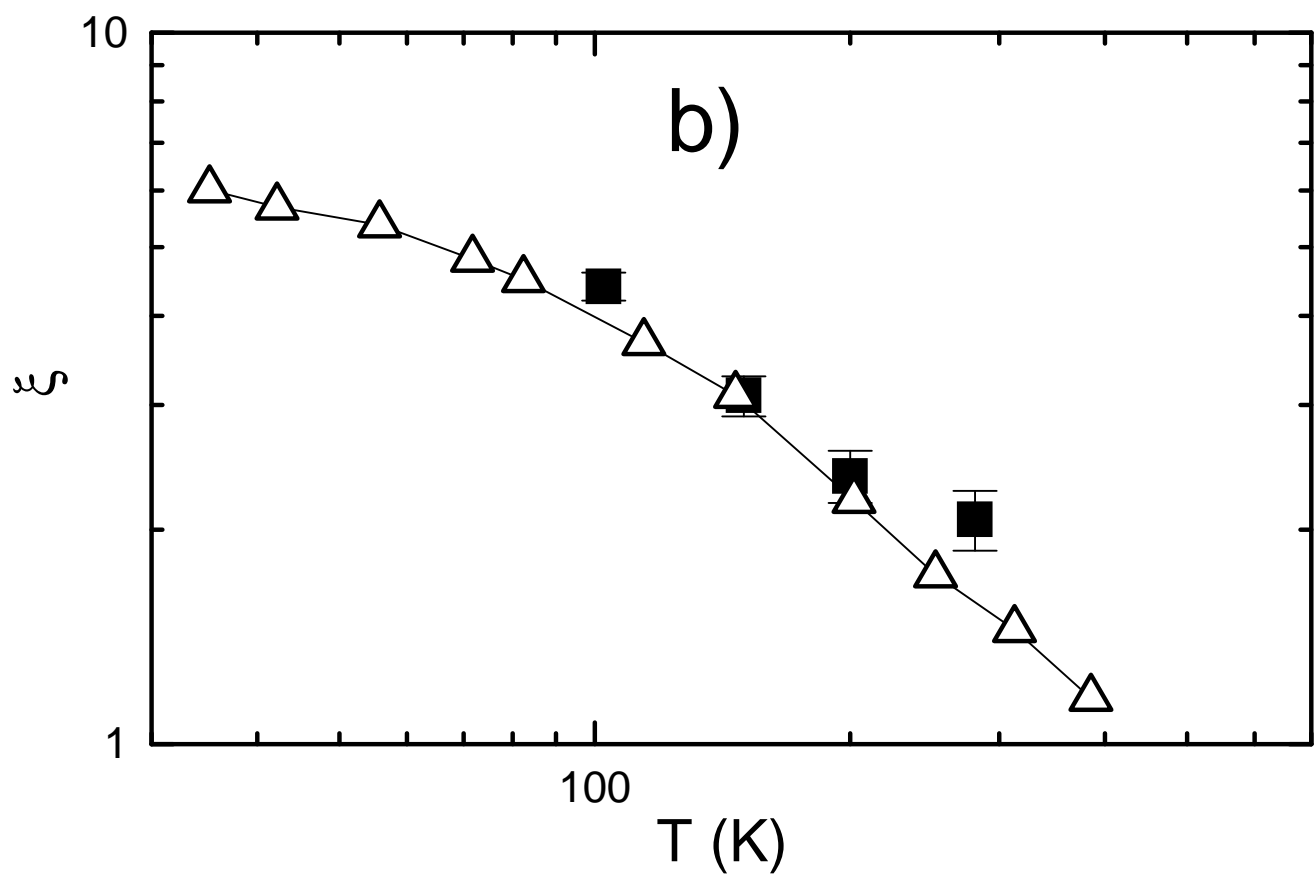
\end{document}